\documentclass[reprint,aps,prb,twocolumn,superscriptaddress,floatfix,showpacs,amsmath,amssymb]{revtex4-1}
\usepackage{verbatim}
\usepackage{graphicx,color,bbm, import}
\usepackage{epsfig}
\usepackage{amsmath}
\usepackage{bbold}
\usepackage{capt-of}
\usepackage{braket}
\usepackage{dirtytalk}
\newcommand{\ketbra}[2]{\vert {#1} \rangle \langle{#2}\vert}

\usepackage[normalem]{ulem}
\begin{document}
\preprint{APS/123-QED}

\title{Non-equilibrium thermodynamics of continuously measured quantum systems: a circuit-QED implementation}
\author{P.G. Di Stefano}
\affiliation{Centre for Theoretical Atomic, Molecular and Optical Physics,
School of Mathematics and Physics, Queen's University Belfast, Belfast BT7 1NN, United Kingdom.}
\affiliation{Dipartimento di Fisica e Astronomia,
Universit\`a di Catania, Via Santa Sofia 64, 95123 Catania, Italy.}
\author{J.J. Alonso}
\affiliation{Department of Physics, Friedrich-Alexander-Universit\"at Erlangen-N\"urnberg, D-91058 Erlangen, Germany.}
\author{E. Lutz}
\affiliation{Department of Physics, Friedrich-Alexander-Universit\"at Erlangen-N\"urnberg, D-91058 Erlangen, Germany.}
\affiliation{Institute for Theoretical Physics I, University of Stuttgart, D-70550 Stuttgart, Germany.}
\author{G. Falci}
\affiliation{Dipartimento di Fisica e Astronomia,
Universit\`a di Catania, Via Santa Sofia 64, 95123 Catania, Italy.}
\affiliation{CNR-IMM  UOS Universit\`a (MATIS), 
Consiglio Nazionale delle Ricerche, Via Santa Sofia 64, 95123 Catania, Italy.}
\affiliation{Istituto Nazionale di Fisica Nucleare, Via Santa Sofia 64, 95123 Catania, Italy.}
\author{M. Paternostro}
\affiliation{Centre for Theoretical Atomic, Molecular and Optical Physics,
School of Mathematics and Physics, Queen's University Belfast, Belfast BT7 1NN, United Kingdom.}

\date{\today}
\begin{abstract}
We propose an operational framework to study the non-equilibrium thermodynamics of a quantum system $S$ that is coupled to a detector $D$ whose state is continuously monitored, allowing to single out individual quantum trajectories of $S$. We focus on detailed fluctuation theorems and characterize the entropy production of the system. We establish fundamental differences with respect to the thermodynamics of unmonitored, unitarily evolved systems. We consider the paradigmatic example of circuit-QED, where superconducting qubits can be coupled to a continuously monitored resonator and show numerical simulations using state-of-the-art experimental parameters.
\end{abstract}
\maketitle
\section{Introduction}
\label{sec:intro}
The origin of dynamical irreversibility and the emergence of the arrow of time from the microscopic laws of quantum mechanics have attracted significant interest in the past few years~\citep{ka:211-campisi-entropy, ka:211-jarzynski-annurev, ka:215-batalhao-paternostro-prl-arrow, Korotkov-arrow}. In particular, recent efforts in the field of non-equilibrium quantum  thermodynamics resulted in the characterization of irreversibility in terms of fluctuation theorems~\citep{ka:211-campisi-rmp} and entropy production~\citep{ka:211-deffner-lutz-prl}.

The standard formulation of non-equilibrium thermodynamic quantities uses explicitly time-gated multi-measurement strategies~\citep{ka:211-campisi-rmp}. Notwithstanding the success encountered by such formulations in describing the thermodynamic implications of non-equilibrium processes all the way down to the quantum domain~\citep{batalhao-paternostro-prl-experiment,An,Cerisola}, such requirements are very difficult to be met in practice. Indeed, the common experimental configurations typically involve the continuous interaction between a system and a measurement apparatus. Such interaction can result in either strong projective measurements inducing ``quantum jumps'' on the state of the quantum system at hand~\citep{E0,E1,E2,E3,E4,E5}, or in the acquisition of only partial information on it. Recently, a theoretical framework for the analysis of stochastic thermodynamics of weakly monitored quantum systems was put forward~\citep{Alonso2016,Elouard}.

In this paper, we make further steps along the lines of defining a fully operational framework for stochastic thermodynamics of continuously monitored systems by considering the case of a dynamical detector coupled to a system of interest and being continuously monitored. This situation adheres perfectly with the configurations typically engineered and encountered in a wide range of experiments. In particular, superconducting circuit quantum electrodynamics (circuit-QED) systems~\citep{Solinas1, Solinas2} embody a very suitable platform, where the system is typically provided by a set of superconducting information carriers, while the field of a stripline resonator plays the role of the continuously monitored dynamical detector~\citep{Blais}. This offers a virtually ideal scenario for the study of stochastic thermodynamics of continuously monitored systems, and the investigation of the deviations from the time-gated approach that has dominated the field to date. In particular, our work sets the theoretical context for the experimental analysis of irreversibility in the non-equilibrium dynamics of a driven superconducting device as quantified by the irreversible entropy production, and the test of the continuous-monitoring version of fundamental fluctuation theorems. 

The remainder of this paper is structured as follows. In Sec.~\ref{sec:closed} we review the non-equilibrium thermodynamics of closed quantum systems, while in Sec.~\ref{sec:cqed} we present the circuit-QED model and the non-equilibrium thermodynamics of continuously monitored circuit-QED systems. In Sec.~\ref{sec:res} we show numerical results for the entropy production and detailed fluctuation theorems. Finally, in Sec.~\ref{sec:con} we draw our concluding remarks.
\section{Non-equilibrium thermodynamics of closed quantum systems}
\label{sec:closed}
The typical setting for a non-equilibrium thermodynamics experiment in closed quantum systems is the following: a system $S$ of Hamiltonian $H_S(\lambda_t)=\sum \epsilon_k(\lambda_t) \ketbra{n^{\lambda_t}}{n^{\lambda_t}}$ is initially (time $t=0$) in equilibrium with its environment at inverse temperature $\beta$, i.e. $\rho_S(0) = \rho_0$, where we defined the Gibbs state
$\rho_t={\text{e}^{-\beta H_S(\lambda_t)}}/{{\cal Z}_t}$
with ${\cal Z}_t= \text{Tr}[\text{e}^{-\beta H_S(t)}]$ the partition function. It is then brought out of equilibrium by the application of an external force protocol $\lambda_t$ parametrized in the time interval $[0,\tau]$. In the closed quantum systems scenario, it is assumed that in $[0,\tau]$ the system is effectively detached from its environment and that $S$ evolves unitarily through the time-evolution operator $U_{t_1,t_2}:={\cal T} \mathrm{e}^{- i \int_{t_1}^{t_2} dt^{\prime}\ H_S(\lambda_{t^{\prime}})}$, where ${\cal T}$ is the time-ordering operator. The non-equilibrium work performed on the system is usually defined~\citep{ka:207-talkner-work} as a stochastic variable $W$ whose single realizations $\epsilon_m(\lambda_{\tau})-\epsilon_n(\lambda_0)$ are weighted by the probability of observing a $\ket{n^{\lambda_0}} \to \ket{m^{\lambda_{\tau}}}$ transition due to the application of the force protocol. Identifying $p(m^{\lambda_{\tau}},n^{\lambda_0}) = \text{Tr}[\Pi_m^{\tau}{\cal U}_{\tau,0}\Pi_n^0\rho_0\Pi_n^0{\cal U}_{\tau,0}^{\dagger}]$, where $\Pi_k^t = \ketbra{k^{\lambda_t}}{k^{\lambda_t}}$, as the probability for such a transition to occur, one may define the work distribution as $p_F(W) = \sum_{m^{\lambda{\tau}},n^{\lambda_0}} p(m^{\lambda_{\tau}},n^{\lambda_0}) \delta(W-\epsilon_m(\lambda_{\tau})+\epsilon_n(\lambda_0))$. In order to address irreversibility, the corresponding \textit{backward} work distribution is usually considered, where the force protocol is reversed in time. One then looks at the probability of the backward transition $\Theta \ket{m^{\lambda_{\tau}}} \to \Theta \ket{n^{\lambda_0}}$, $\Theta$ being the time-reversal operator, with initial statistics given by the Gibbs state $\tilde{\rho}_{\tau}=\Theta \rho_{\tau} \Theta^{\dagger}$ at time $ t= \tau$ when the backwards protocol $\tilde{\lambda}_t = \lambda_{\tau-t}$ is applied. In considering the backwards protocol, we will assume that the Hamiltonian of the systems obeys a time-reversal symmetry of the form $\Theta H_S(\lambda_t) \Theta^{\dagger} = \epsilon_{\lambda} H_S(\lambda_{\tau-t})$ where $\epsilon_{\lambda} = \pm 1$ \citep{ka:210-campisi-philtransrsoc}. We call $p_B(W)$ the corresponding backward work distribution and state the Crooks fluctuation theorem~\citep{ka:198-crooks-pre}
\begin{equation}
\label{eq:crooks}
{p_F(W)}/{p_B(-W)}=\mathrm{e}^{\beta(W-\Delta F)},
\end{equation}
where we used the free energy difference $\Delta F = -(1/\beta) \log({\cal Z}_{\tau}/{\cal Z}_0)$. By integrating over $W$ one gets the celebrated Jarzynski identity $\langle \text{e}^{-\beta(W-\Delta F)}\rangle=1$, which entails the second law through the Jensen inequality $\langle \Sigma \rangle \geq 0$, where the irreversible entropy production $\Sigma := \beta(W-\Delta F)$ has been defined~\citep{Jarzynski-prl}.

The closed quantum systems paradigm is in contrast with the approach of classical stochastic thermodynamics. In the latter, work realizations are described in terms of trajectories of a classical system in phase space. In this paper, we propose an implementation of a non-equilibrium thermodynamics experiment using the framework of quantum \textit{stochastic} thermodynamics~\citep{Alonso2016,Elouard}. We thus exploit the formalism of quantum trajectories considering a system that is continuously monitored during its evolution through the coupling with a detector $D$. By doing so, we are able to single out individual quantum trajectories and characterize irreversibility in a way that is compatible with the classical picture. Despite methodological similarities, though, we point out that important differences arise due to the back-action of quantum measurement on the system state.

\begin{figure}
\includegraphics[width=\linewidth]{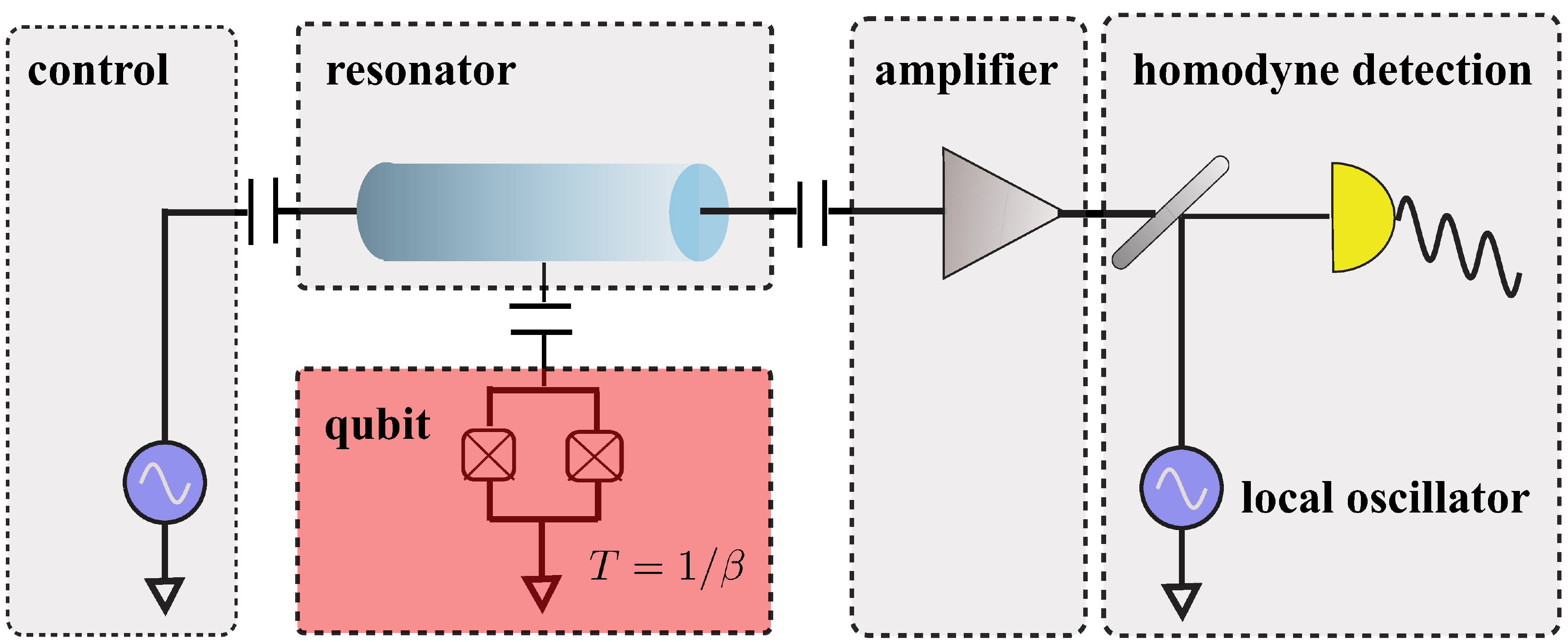}
\caption{(Color online) Setup for a circuit QED implementation. A superconducting qubit is coupled to a resonator, through which it is measured and controlled. Measurement is performed by means of continuous homodyne observation of the amplified cavity field. A strong local oscillator provides a second amplification stage producing an output current $I(t)$ encoding information about qubit and resonator.\label{fig:setup}}
\end{figure} 
\section{Quantum thermodynamics of continuously monitored systems: the Circuit-QED case}
\label{sec:cqed}
We address the typical platform of circuit-QED implementations, such as the one depicted in Fig.~\ref{fig:setup}. We thus consider a superconducting qubit, e.g. a transmon~\citep{ka:207-koch-pra-transmon}, coupled to a microwave resonator in the strong dispersive coupling regime. The latter is used both to drive (thus acting as a forcing mechanism) and measure the qubit~\citep{Blais}. Recently, continuous monitoring in circuit-QED has been successfully employed to observe single quantum trajectories of a transmon qubit~\citep{ka:213-murchsiddiqi-nature,ka:214-roch-prl,ka:212-vijay-siddiqi-nature-feedback,ka:213-hartridge-devoret-science-backaction} and quantum jumps~\citep{ka:211-vijai-siddiqi-prl-jumps,Naghiloo,Cottet,Masuyama,Naghiloo2}. We shall regard the qubit as our system of interest $S$, and the resonator as the detector $D$. Together, system and detector are driven by a ``forcing" field oscillating at frequency $\omega(t)/2\pi$ and almost resonant with the qubit frequency $\omega_0/2\pi$, and by a ``measurement" field having angular frequency $\omega_d$ that is almost resonant with the resonator frequency $\omega_c/2\pi$. Qubit and resonator will be dispersively coupled, i.e. $g \ll \Delta$, where $g$ is the strength of the coupling and $\Delta=\omega_0-\omega_c$ is the cavity-qubit detuning. We will also assume $g \ll \omega_{0,c}$, i.e. we will be outside  the so-called ultra-strong coupling regime\citep{ka:210-niemczyck-natphys-ultrastrong}, as it is the case in most of the implementations reported so far in the literature. The Hamiltonian of the total system can be split as
\begin{equation}
\label{eq:hamiltonian}
H = H_S + H_D + H_{\text{int}}
\end{equation}
where we have introduced the detector Hamiltonian $H_D = \omega_0\ a^{\dagger}a + \epsilon_d\ (a \text{e}^{i\omega_d t} + a^{\dagger}  \text{e}^{-i\omega_d t})$ and the $S-D$ interaction Hamiltonian is $H_{\text{int}} = \chi\sigma_z a^{\dagger}a$. Here, $a$ is the annihilation operator for the resonator field and $\sigma_i$ are the usual Pauli operators. Here $\chi= g^2/\Delta$ is an effective coupling in the dispersive regime determining a Stark shift of the cavity frequency conditioned to the qubit state, which is the physical mechanism for the qubit detection. The dispersive regime also implies weak coupling between qubit and field, allowing us to separately define energies. The system Hamiltonian $H_S$ can be split into
\begin{equation}
\label{eq:hs}
H_S = H_0 + H_{\lambda_t},
\end{equation}
where $H_0 = \omega_0 \sigma_z/2$ is the bare Hamiltonian of the qubit and $H_{\lambda_t}=\delta \omega_0(t) \sigma_z/2 + \Omega(t)\cos (\varphi(t))\sigma_x$ is the time-dependent contribution that implements the force protocol. In circuit-QED, the available control that can be exploited in order to manipulate the system breaks down into independent tunability of both the qubit frequency $\delta \omega_0(t)$, achieved through the application of a time-dependent magnetic field in the SQUID loop of the transmon, and the parameters of the external microwave field, i.e. the amplitude $\Omega(t)$ and phase $\varphi(t)$.

The conditional Stark shift $H_{\text{int}}$ allows for the state of the qubit to be mapped onto a quadrature of the field, which we define as $X_{\phi} = (a \text{e}^{i\phi}+a^{\dagger} \text{e}^{-i\phi})/\sqrt{2}$ with $\phi\in[0,2\pi]$ a phase. In our model, the qubit state is mapped onto the in-phase quadrature $X_0$. Continuous monitoring can thus be done through homodyne measurements of the field leaking out of the resonator at rate $\kappa$~\citep{ka:208-gambetta-pra-trajectory}. The homodyne photocurrent resulting from the mixing of the cavity field with a strong local oscillator tuned on the phase of the quadrature $X_0$ is continuously observed, thus inducing quantum back-action on the $S-D$ system. The evolution of the latter over a single quantum trajectory will be thus conditional on the measured photocurrent. In order to describe the dynamics of the system, we partition the time interval $[0,\tau]$ into small but finite time intervals $\delta t=t_{i+1}-t_i~(i=0,..,N)$ with $t_0=0$ and $t_{N+1}= \tau$. Here, $\delta t$ is chosen to be much smaller than the shortest time-scale of the problem, so that we can approximate ${\cal U}_{t_i,t_{i+1}} \simeq \mathbb{1} -i H \delta t$. 

The effect of a measurement can be modelled through the positive operator valued measurement (POVM) $L_{x}$, such that $\int dx\ L_{x}^{\dagger} L_{x} = \mathbb{1}$, where $x$ refers to the average value of the homodyne photocurrent over $\delta t$. In the small time interval $\delta t$, the overall dynamics of the system can be effectively factorized into two independent contributions given by unitary evolution and measurement. By introducing the operators ${\cal O}_{t_i}= L_{I(t_{i+1})} {\cal U}_{t_i,t_{i+1}}$, the evolution of the system, conditional to the observation of the stream of average photocurrents $I= \{I(0), I(t_1), ... , I(t_k)\}$, is thus given by
\begin{equation}
\rho_{D+S}(t_k) = \frac{(\overset{\leftarrow}{\prod}_{i<k} {\cal O}_{t_i} )\rho_{D+S}(0)(\overset{\rightarrow}{\prod}_{i<k} {\cal O}_{t_i}^{\dagger})}{\text{Tr}[(\overset{\leftarrow}{\prod}_{i<k} {\cal O}_{t_i} )\rho_{D+S}(0)(\overset{\rightarrow}{\prod}_{i<k} {\cal O}_{t_i}^{\dagger})]},
\end{equation} 
where the arrows imply time ordering. In the homodyne measurement scheme for circuit-QED, measurement operators are given by $L_{x}=[1 -\kappa a^{\dagger}a\ \delta t/2 + x \sqrt{\kappa} \delta t] \sqrt{p_o(x)}$, where $p_o(x)= \exp(-\delta t x^2)/\sqrt{\delta t/2\pi}$ is the ostensible~\citep{kb:214-wiseman-milburn} probability density of obtaining the result $x$ for the homodyne photocurrent. We should point out that additional decoherence terms may add up in the dynamics of the system, caused by relaxation and dephasing of the qubit. We did not include those terms in our analysis since, as it will be argued later, decoherence rates are small enough in present technology to have a negligible effect in the time scale relevant to the experiment.

The statistics of the qubit alone is, in general, given by partial tracing over the detector degrees of freedom. Nonetheless, in the limit of a sufficiently weak measurement, i.e. when the average number of photon is $\bar{n}=(\epsilon_d/\kappa)^2 \ll 1$, $\chi \ll \kappa$ and the driving is weak, i.e. $\Omega \ll \kappa$, the qubit and the detector develop negligible entanglement~\citep{Korotkov2011} and the dynamics of $S$ can be factorized from the dynamics of $D$. The qubit density matrix at time $t_k$ will be therefore given by
\begin{equation}
\label{eq:stochastic-qubit}
\rho_{S}(t_k) = \frac{(\overset{\leftarrow}{\prod}_{i<k} {\cal Q}_{t_i} )\rho_S(0)(\overset{\rightarrow}{\prod}_{i<k} {\cal Q}_{t_i}^{\dagger})}{\text{Tr}[(\overset{\leftarrow}{\prod}_{i<k} {\cal Q}_{t_i} )\rho_S(0)(\overset{\rightarrow}{\prod}_{i<k} {\cal Q}_{t_i}^{\dagger})]}
\end{equation}
where ${\cal Q}_{t_i} = M_{I(t_{i+1})} \text{e}^{-i \int_{t_i}^{t_{i+1}} dt H_S(t)}$ and the POVM operators for the qubit alone are given by
\begin{equation}
\label{eq:measurement-qubit}
M_{x} = \sqrt{P_0(x)} \ketbra{0}{0} + \sqrt{P_1(x)} \ketbra{1}{1}
\end{equation}
Here we defined the probability distributions $P_j(x) = \text{e}^{-i \delta t/2 (x + (-1)^j \sqrt{\Gamma_d})^2}$~\citep{ka:216-feng-scirep-exact-bayes}(cf. Appendix~\ref{sec:app}), with the measurement rate given by $\Gamma_d = 16 \chi^2\bar{n}/\kappa$.

During its evolution, the system experiences hanges in its internal energy $U(t)= \text{Tr}[\rho_S H_S]$. The infinitesimal variation $dU(t_i) = \delta W_i + \delta Q_i$ of the latter can be split into a unitary and a back-action term, i.e.
\begin{equation}
\delta W_i = \text{Tr}[\rho_S(t_i) \ d H_S(t_i)],~\delta Q_i = \text{Tr}[H_S(t_i) \ d \rho_S(t_i)]
\end{equation}
where the discretized differential is $d X(t_i) = X(t_{i+1})-X(t_i)$. The term dubbed $\delta W_i$ clearly embodies a contribution to work, as it quantifies the average change of Hamiltonian of the system. On the other hand, the term $\delta Q_i$ is identically null whenever the system evolves via a unitary (i.e. Hamiltonian) dynamics, and can thus be associated with the non-unitary contribution to the change of internal energy, that is heat. Correspondingly, we will define work and heat as $W(t) = \sum_{t_i<t} \delta W_i$ and $Q(t) = \sum_{t_i<t} \delta Q_i$ respectively. Notice that the above definition of work is fundamentally different from the usual one for closed systems. In the latter case, as mentioned above, work realizations are determined as differences between eigenvalues of the final and initial Hamiltonian. For a continuously monitored system, on the other hand, work can be defined at the single trajectory level as a time-dependent stochastic process. This is similar, in spirit, to the approach of classical stochastic thermodynamics~\citep{ka:212-seifert-review} with the fundamental difference that, while in classical physics a trajectory in phase space can be monitored without disturbing its dynamics, measurement back action plays a fundamental role in quantum systems, generating the heat term $Q$. The latter has been given a straightforward interpretation in Refs.~[\onlinecite{Alonso2016,Naghiloo}] as the amount of work necessary to isolate the system using quantum feedback, or alternatively in Ref.~[\onlinecite{Elouard}] as the amount of work an external daemon would need to contribute in order to counter quantum back-action.

We are now concerned with the characterization of irreversibility for this system, which can be done by means of detailed fluctuation theorems. The probability of observing a particular single trajectory in the Hilbert space, though, cannot be defined, as it was the case for closed systems, only through its end points. The stochastic evolution of the system's state have, in fact, to be taken into account. From Eq.~(\ref{eq:stochastic-qubit}), we see that a single trajectory can be fully characterized if we also take into account the measured current $I(t)$. The probability of observing a trajectory starting from $\ket{n^{\lambda_0}}$ and ending in $\ket{m^{\lambda_{\tau}}}$ while measuring  $I(t)$ is then given by 
\begin{equation}
\label{eq:forward-probability}
\begin{aligned}
&p_F(m^{\lambda_{\tau}},I(t), n^{\lambda_0}) = \text{Tr}[\Pi_m^{\tau} (\prod^{\leftarrow}_i {\cal Q}_{t_i} )\Pi_n^0\rho_0\Pi_n^0(\prod^{\rightarrow}_i {\cal Q}_{t_i}^{\dagger} )]
\end{aligned}
\end{equation}
\begin{figure}[t]
\begin{minipage}{\linewidth}
\includegraphics[width=\linewidth]{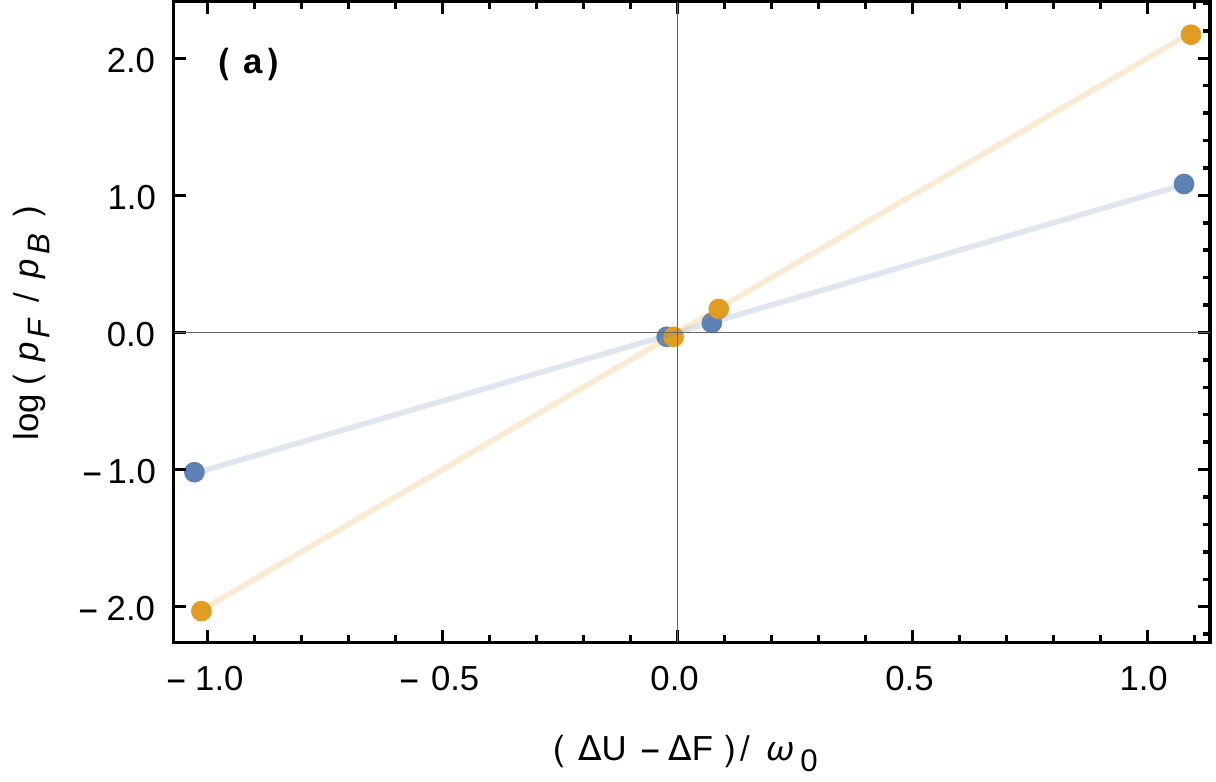}
\end{minipage}
\begin{minipage}{\linewidth}
\includegraphics[width=\linewidth]{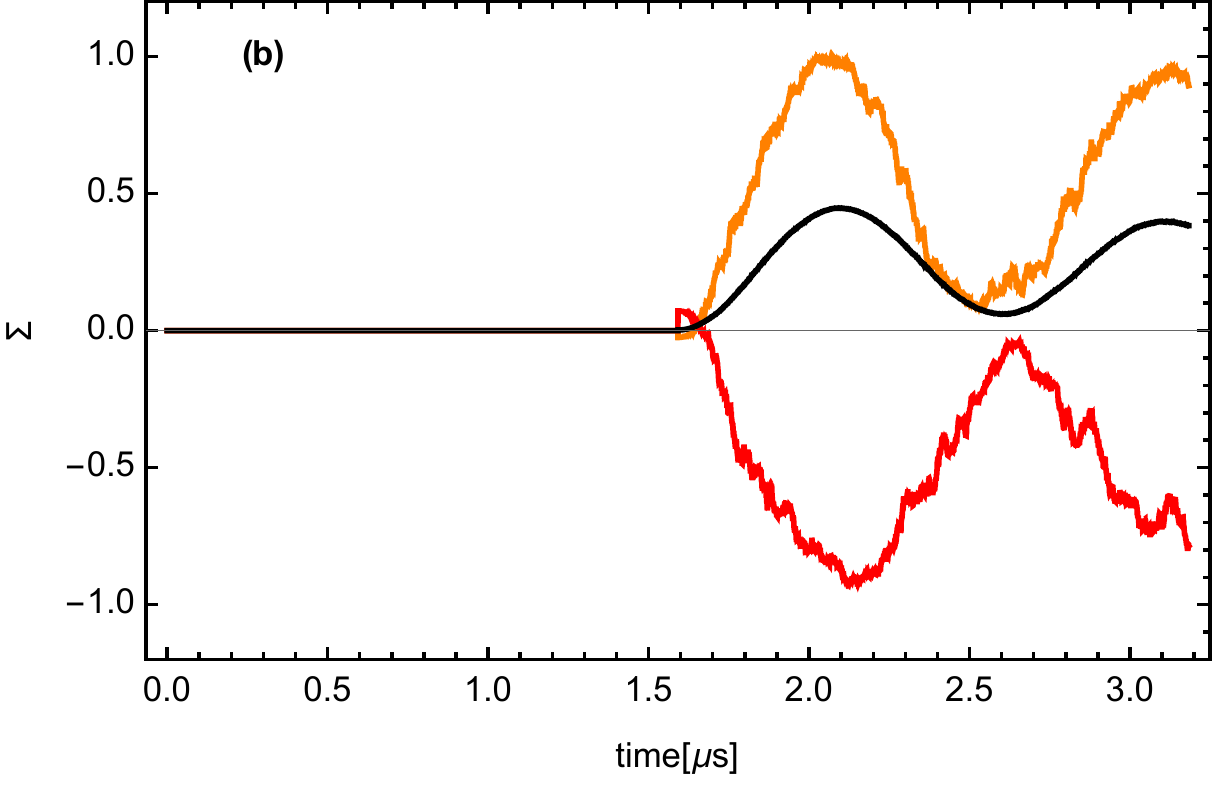}
\end{minipage}
\caption{(Color online) Numerical results for the protocol of Eq.~(\ref{eq:force-protocol}). {Panel (a)}: Logarithmic representation of the detailed fluctuation theorem of Eq.~\eqref{eq:watanabe-crooks}. Here we used $\beta = 1/\omega_0$ (blue curve) and $\beta=\/\omega_0$. {Panel (b)}: Entropy production of two trajectories (red and orange curves) for $\beta = 1/\omega_0$ showing how negative entropy production trajectories can observed. The mean entropy production (black curve), though, is always non negative.\label{fig:results}
}
\end{figure}
In the spirit of the detailed fluctuation theorem of Eq.~(\ref{eq:crooks}), we will again consider a backwards trajectory starting in $\Theta \ket{m^{\lambda_{\tau}}}$ and ending in $\Theta \ket{n^{\lambda_0}}$ where the time-reversal force protocol is applied together with the time-reversal POVM operators $\tilde{M}_{x}$. Employing the operators $\tilde{\cal Q}_{t_i} = {\cal U}_{t_i,t_{i+1}}[\tilde{\lambda}(t)] \tilde{M}_{\tilde{I}(t_{i+1})}$, where $\tilde{M}_{\tilde{I}(t)} = \theta M^{\dagger}_{I(\tau-t)} \theta^{\dagger}
$ are the time-reversed measurement operators of the current $\tilde{I}(t)$ detected in the backward process, we define the probability of the backward trajectory as
\begin{equation}
\begin{aligned}
&p_B(n^{\lambda_0},\tilde{I}(t), m^{\lambda_{\tau}}) = \text{Tr}[\tilde{\Pi}_n^{0}(\prod^{\leftarrow}_i \tilde{{\cal Q}}_{t_i} )\tilde{\Pi}_m^{\tau}\tilde{\rho}_{\tau}\tilde{\Pi}_m^{\tau}(\prod^{\rightarrow}_i \tilde{{\cal O}}_{t_i}^{\dagger} )],
\end{aligned}
\end{equation}
where $\tilde{\Pi}_k = \Theta \Pi_k \Theta^{\dagger}$ and $\tilde{\rho}_{\tau} = \Theta \rho_{\tau} \Theta^{\dagger}$. It has been shown in Ref.~[\onlinecite{Watanabe2014}] that, under projective measurements, one can state a generalized detailed fluctuation theorem as
\begin{equation}
\label{eq:watanabe-crooks}
\frac{p_F(m^{\lambda_{\tau}},I(t), n^{\lambda_0})}{p_B(n^{\lambda_0},\tilde{I}(t), m^{\lambda_{\tau}})} = \text{e}^{\beta(\Delta U_{nm} - \Delta F)},
\end{equation}
where the internal energy difference is defined as $\Delta U_{nm} = \epsilon_m(\lambda_{\tau}) -\epsilon_n(\lambda_0)$. We should point out that, in general, with such a notion of time reversal, $p_B$ is not a proper probability distribution, i.e. $\int  {\cal D}I \  p_B(n^{\lambda_0},\tilde{I}(t), m^{\lambda_{\tau}}) \neq 1$, where ${\cal D}I$ is a measure for the path integral. The normalization condition holds if and only if $\int \text{d}x\  M_xM^{\dagger}_x = \mathbb{1}$, which is indeed the case for the reduced dynamics for the qubit, where the measurements operators are hermitian. The probability distribution of the internal energy can then be written as
\begin{equation}
p(\Delta U) = \sum_{m,n} \int {\cal D} I\ p_F(m^{\lambda_{\tau}},I(t), n^{\lambda_0}) \delta(\Delta U - \Delta U_{nm}),
\end{equation}
which immediately gives
\begin{equation}
\label{eq:jarzinsky-cont-mon}
\langle \text{e}^{- \Sigma} \rangle = 1.
\end{equation}
Here, the entropy production $\Sigma = \beta (W + Q - \Delta F)$ allows again to derive the second law of thermodynamics $\langle \Sigma \rangle \geq 0$ through the use of Jensen inequality, while the quantity $\langle \text{e}^{- \Sigma} \rangle$ is customarily termed {\it efficacy}. Notice that the heat term~\citep{Elouard} is a unique feature of quantum back-action, and has thus no equivalent neither in the closed quantum system case, nor in the classical stochastic thermodynamics case. 

\section{Numerical Results}
\label{sec:res}
In order to numerically simulate our analysis, we made use of parameters borrowed from state of the art technology~\citep{ka:213-murchsiddiqi-nature}. In particular, we considered an architecture involving a transmon with $\omega_0/2\pi = 4\ \mathrm{GHz}$, a leaking rate of the resonator at $\kappa/2\pi = 10\ \mathrm{MHz}$ and a coupling constant $\chi/2\pi = -0.5\ \mathrm{MHz}$.
\begin{figure}[t]
\begin{minipage}{\linewidth}
\includegraphics[width=\linewidth]{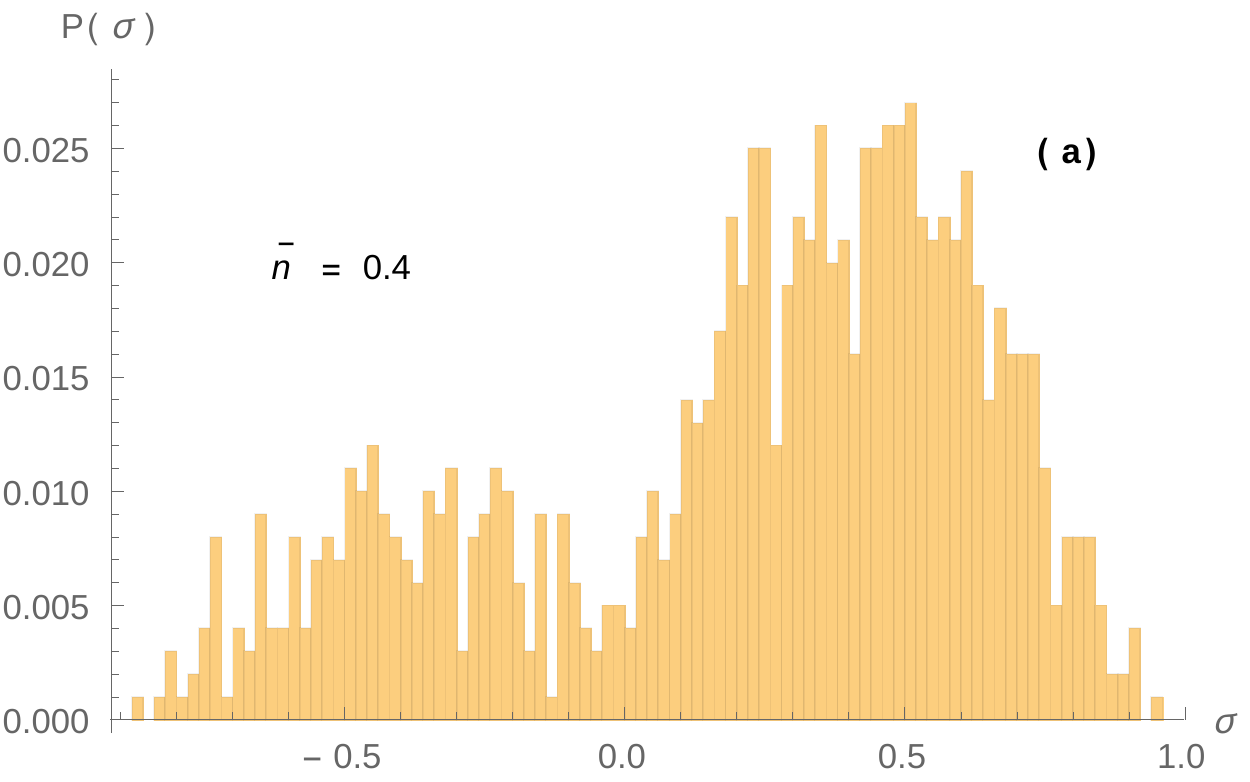}
\end{minipage}
\begin{minipage}{\linewidth}
\includegraphics[width=\linewidth]{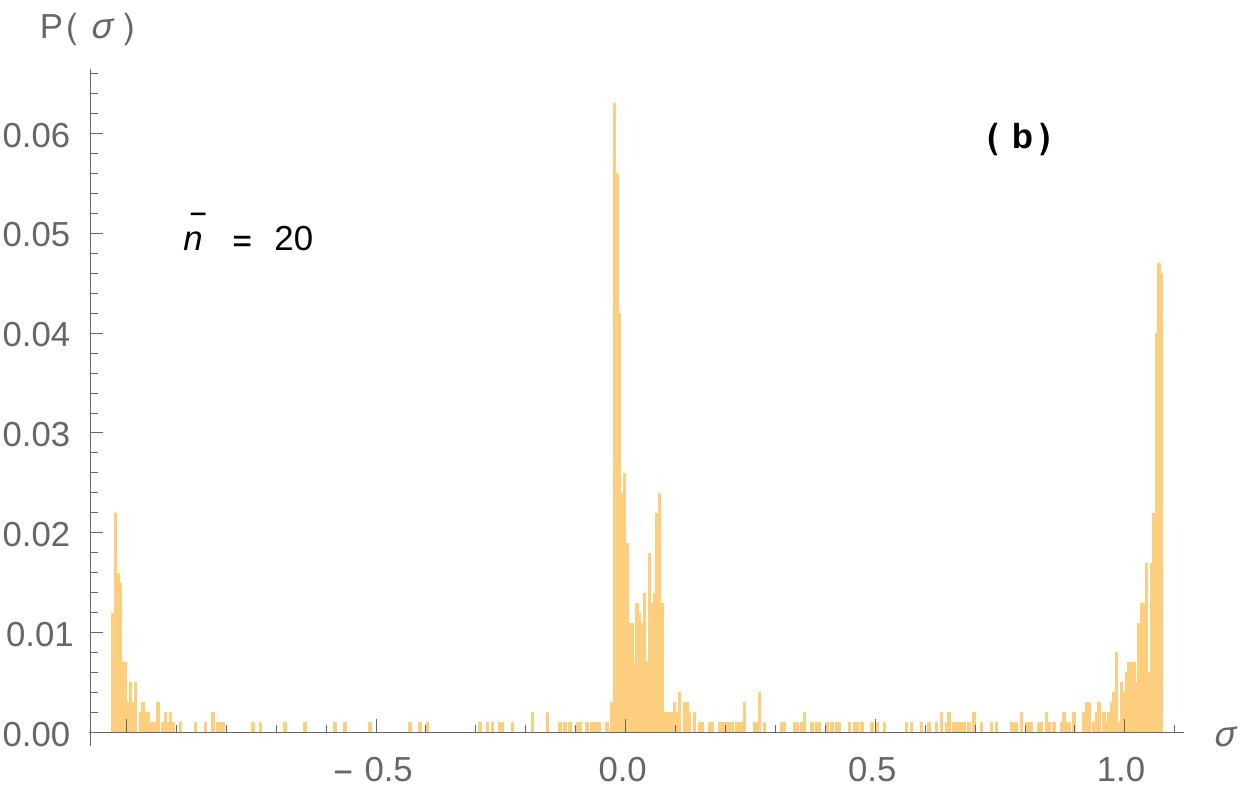}
\end{minipage}
\begin{minipage}{\linewidth}
\includegraphics[width=\linewidth]{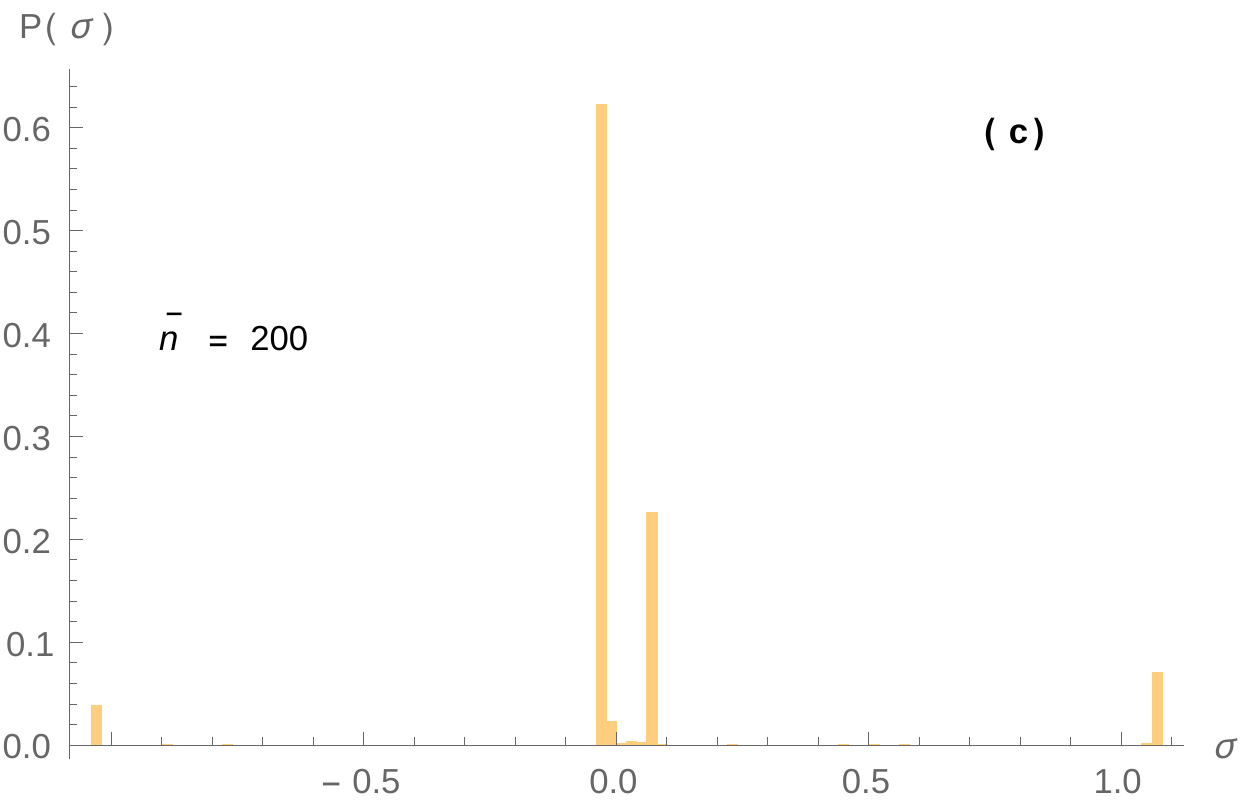}
\end{minipage}
\caption{(Color online) Distribution of the entropy production $\Sigma = \beta (W+Q - \Delta F)$ at $t = 2.4 \mathrm{\mu s}$ along $10^3$ Monte Carlo trajectories for an average number of photons $\bar{n} = 0.4$ [panel (a)], $\bar{n} = 20$ [panel (b)], and $\bar{n} = 200$ [panel (c)].\label{fig:results-hist}
}
\end{figure}

We considered a force protocol in which the frequency of the qubit and the amplitude of the field are both quenched, i.e.
\begin{equation}
\label{eq:force-protocol}
\delta \omega_0(t) = \Delta \omega\, \theta \left( t-\frac{\tau}{2} \right), \quad \Omega(t) = \Omega_0\, \theta\left( t-\frac{\tau}{2} \right),
\end{equation} 
where we used values $\Delta \omega/2\pi = 400 \mathrm{MHz}$ and $\Omega_0/2\pi = 1 \mathrm{MHz}$, while we kept the external drive frequency constant, i.e. $\dot{\varphi}(t) =: \omega = \omega_0 + \Delta \omega$.

For the simulations of Fig.~\ref{fig:results} we set the amplitude of the measurement field so to have, as in Ref.~[\onlinecite{ka:213-murchsiddiqi-nature}], an average number $\bar{n}=2(\epsilon_d/\kappa)^2 = 0.4$ of photons in the cavity. These parameters yield a measurement rate $\Gamma_d/2\pi = 160 \mathrm{KHz}$, which gives a measurement time~\citep{ka:208-gambetta-pra-trajectory} $t_m = 1/(2\Gamma_d) \simeq 500 \mathrm{ns}$, which is usually much smaller than energy relaxation and pure dephasing times for a state-of-the-art transmon ($T_1\sim T_2^{\ast} \gtrsim 10 \mathrm{\mu s}$). This enabled us to neglect energy relaxation and dephasing in our model. Figure~\ref{fig:results}{(a)} shows how the detailed fluctuation theorem of Eq.~(\ref{eq:watanabe-crooks}) can be tested. In particular, we plot the equation $\log p_F/p_B = -\beta (\epsilon_m(\lambda_{\tau})-\epsilon_n(0) - \Delta F)$, which is verified by showing how the slope of the interpolation lines in Fig.~\ref{fig:results}{(a)} equals $\beta$. In particular, we employed values $\beta = 1/\omega_0$ (blue curve) and $\beta = 2/\omega_0$ (orange curve). In Fig.~\ref{fig:results}{(b)} we show trajectories for the entropy production (colored lines) together with the mean entropy production (black line). We notice how, while the mean entropy production is non-negative as required by the second law of thermodynamics, stochastic thermodynamics attained through continuous monitoring of a small quantum system, such as a transmon, allows to observe \textit{negative} entropy production trajectories, in striking contrast with what one experiences in the macroscopic world.

In Figs.~\ref{fig:results-hist}(a)-(c) we show the distribution of the entropy production at $t = 2.4 \mathrm{\mu s}$ over $10^3$ Monte Carlo trajectories for an increasing average number of photons $\bar{n}$ in the cavity, i.e. for a growing measurement strength. We notice how the distributions show multiple components, the bigger ones being in the positive semi-axis. By increasing the measurement strength, the distribution of entropy production develop mutually isolated peaks, displaying a clear multi-modal character. While for $\bar{n} = 0.4$ [Fig.~\ref{fig:results-hist}(a)] we have  a blurred bi-modal distribution, increasing $\bar{n}$ up to 200 [cf. Fig.~\ref{fig:results-hist}(c)] we get a four-peak distribution. In fact, an increase in the measurement strength corresponds to a change of the the system's dynamics from a diffusive regime (associated with weak measurements) to a quantum-jump one, typical of a strong measurement condition, where coherences are suppressed and the dynamics of the system effectively consists of transitions between energy eigenstates. Therefore, the entropy production assumes the only four possible values allowed in the usual two-measurement process typically used in order to assess the thermodynamics of closed quantum systems.

\begin{figure}[t]
\includegraphics[width=\linewidth]{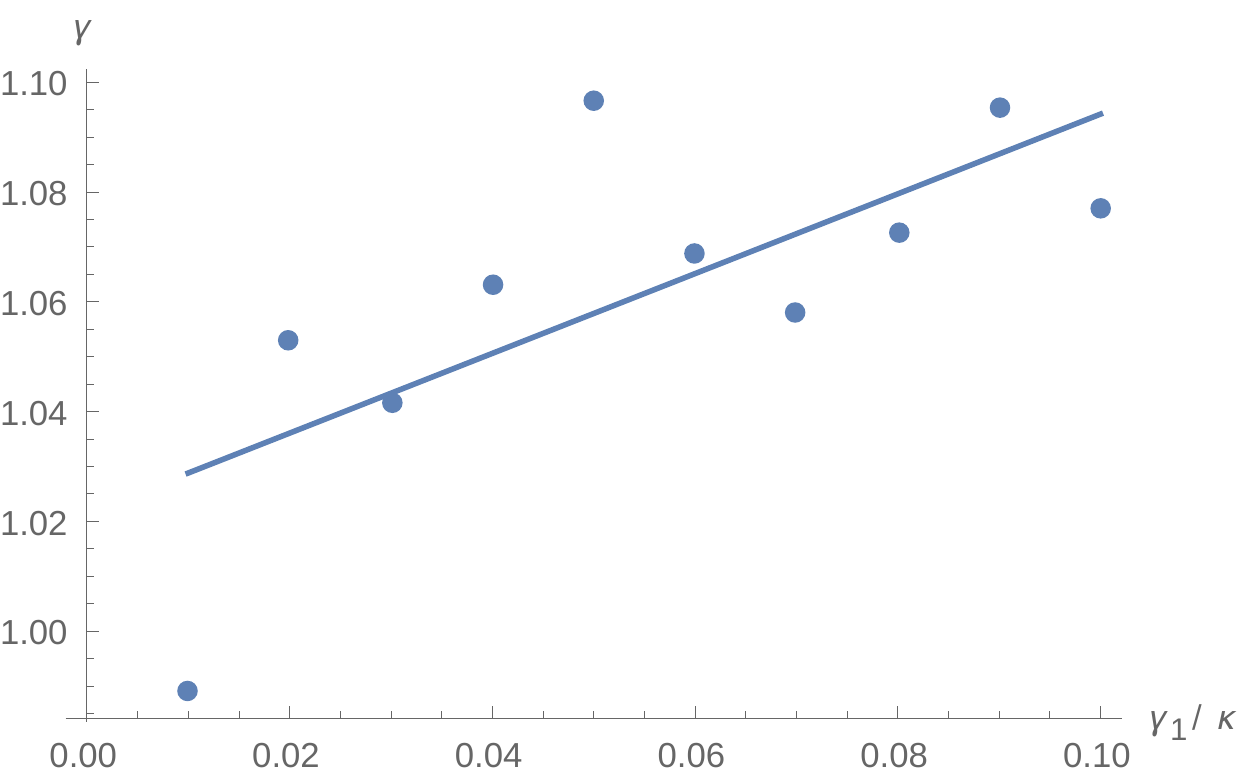}
\caption{(Color online) The points show the efficacy $\gamma = \langle \mathrm{e}^{- \Sigma} \rangle$, in the presence of a dissipative environment, evaluated using $10^3$ Monte Carlo trajectories per data point, for increasing values of the damping rate $\gamma_1$ in Eq.~\eqref{eq:lindblad}. Here we used $\bar{n} = 0.4$. The regression line is ${\gamma} = a + b \gamma_1/\kappa$, with $a = 1.02$ and $b = 0.73$. Here, $b$ has a $p$-value $p < 0.05$, showing a significant relation between $\gamma$ and $\gamma_1/\kappa$.\label{fig:results-gamma}
}
\end{figure}

Finally, we allow for the interaction of the transmon qubit with a dissipative environment. This can be done by modifying our formalism to include for a non-unitary evolution between consecutive measurements. 
We model such non-unitary evolution by assuming that, within the time intervals $\delta t,$ the qubit evolves according to the Lindblad Master Equation 
\begin{equation}
\label{eq:lindblad}
\dot{\rho}_S =  -i[H_S,\rho_S] + \gamma_1\ \sigma_- \rho_S \sigma_+ -\gamma_1\ \frac{1}{2}\{ \sigma_+ \sigma_-, \rho_S \},
\end{equation}
where $\gamma_1$ is the rate of dissipation within such time intervals. In Fig.~\ref{fig:results-gamma} we show how the result of Eq.~(\ref{eq:watanabe-crooks}), and thus Eq.~(\ref{eq:jarzinsky-cont-mon}), break down under such dynamical assumptions. We showcase the behavior of the efficacy $\gamma = \langle \mathrm{e}^{ - \Sigma} \rangle$~(cf. Ref.~[\onlinecite{ka:212-dvir}]) against $\gamma_1$. The noise in the data points is due to the stochasticity of the Monte Carlo trajectories, but a linear regression over the data points shows a significant trend. A non-unit efficacy in a dissipative system results from the fact that the definition of entropy production given above is no longer valid, and should be adapted to include the effects of ``classical'' heat~\citep{Elouard} flowing between the qubit and the environment.
\section{Conclusions}
\label{sec:con}
In this paper we proposed a fully operational framework for the exploration of stochastic quantum thermodynamics resulting from the continuous monitoring of a quantum system. We have considered an ``indirect inference'' case where the system is coupled to a detector, which is continuously monitored, a situation that matches closely a number of experimentally relevant situations. We have shown how the verification of detailed fluctuation theorems and the measurement of witnesses of irreversibility such as the entropy production are easily accessed in circuit-QED architectures. In our simulations, we have shown how high-quality experiments can be set up in existing labs employing state of the art technology. Finally, we have analyzed the role of increasing measurement strengths and decoherence in the entropy production and the fluctuation theorems.

\acknowledgements
This work was supported by the EU Collaborative Projects TherMiQ (grant agreement 618074) and TEQ (grant agreement 766900), and the DfE-SFI Investigator Programme (grant 15/IA/2864). We acknowledge partial support from COST MP1209. 


\appendix
\section{Derivation of the dynamical equations}
\label{sec:app}
In Ref.~[\onlinecite{ka:208-gambetta-pra-trajectory}] a Stochastic Master Equation (SME) was put forward to describe the dynamics of a qubit subject to continuous homodyne measurements via a probing field. Here we will assume that the phase of the local oscillator is tuned to the phase of the quadrature where information on the qubit is encoded~\citep{ka:208-gambetta-pra-trajectory}. With our gauge choice the latter is the in-phase quadrature $X_0$, therefore we will take the phase of the local oscillator to be zero. We shall recast the SME of Ref.~[\onlinecite{ka:208-gambetta-pra-trajectory}] in the form of a Bloch equation in the basis of the bare qubit Hamiltonian $H_0 = \omega_0 \sigma_z/2$. Here, in order to simplify the notation, we shall refer to the system state as $\rho$ and to its Hamiltonian as $H = \delta \omega \sigma_z/2+ \Omega(t) \sigma_x$. The Bloch equations read
\begin{widetext}
\begin{equation}
\label{eq:gambetta-sde}
\begin{aligned}
\dot{\rho}_{00}(t) &= - \dot{\rho}_{11}(t) = -2 H_{01}\ \text{Im}\{\rho_{01}\} + 2 \sqrt{\Gamma_d} \rho_{00}\rho_{11} (I_h - \sqrt{\Gamma_d} \langle \sigma_z \rangle),\\
\dot{\rho}_{01} &= \dot{\rho}_{10}^{\ast} = i H_{00} \rho_{01} + i H_{01} (\rho_{00} - \rho_{11}) -  \sqrt{\Gamma_d}(\rho_{00}-\rho_{11})(I_h - \sqrt{\Gamma_d} \langle \sigma_z \rangle) - \frac{\Gamma_d}{2} \rho_{01},
\end{aligned}
\end{equation}
\end{widetext}
where the full-spectrum homodyne current is $I_h = \sqrt{\Gamma_d}\langle \sigma_z \rangle + \xi (t)$ and $\xi (t)$ is an uncorrelated white Gaussian noise term such that $\text{E}[\xi (t)] = 0, \text{E}[\xi (t_1)\xi (t_2)] = \delta(t_1-t_2)$. For our numerical simulations and in order to take into account the finite bandwidth of the electronics in the circuit, a discretized version of Eq. (\ref{eq:gambetta-sde}) must be employed, reading
\begin{widetext}
\begin{equation}
\label{eq:gambetta-component}
\begin{aligned}
\rho_{00}(t + \delta t) &= \rho_{00}(t) + (-2 H_{01}\ \text{Im}\{\rho_{01}\} + 2 \sqrt{\Gamma_d} \rho_{00}\rho_{11} (I - \sqrt{\Gamma_d} \langle \sigma_z \rangle))\delta t,\\
\rho_{01}(t + \delta t) &= \rho_{01}(t) + (i H_{00} \rho_{01} + i H_{01} (\rho_00 - \rho_{11}) - \sqrt{\Gamma_d}(\rho_{00}-\rho_{11})(I - \sqrt{\Gamma_d} \langle \sigma_z \rangle) - \frac{\Gamma_d}{2} \rho_{01}) \delta t.
\end{aligned}
\end{equation}
\end{widetext}
Here, $\delta t$ is a small but finite time interval such that $\delta t \ll 1/H_{01}, \/\Gamma_d$ and we have introduced the current $I = \frac{1}{\delta t} \int_{t}^{t+\delta t} dt^{\prime}\ I_h(t^{\prime})$. Notice that $I$ can be written as $I = \langle \sigma_z \rangle + \bar{\xi}$, with the stochastic variable $\bar{\xi}$ being distributed following a Gaussian with standard deviation $1/\sqrt{\delta t}$. The probability distribution for $I$ will  then be
\begin{equation}
\label{eq:pi}
P(I) = \sqrt{\frac{\delta t}{2 \pi}} \text{e}^{-\frac{\delta t}{2}(I - \sqrt{\Gamma_d}\langle \sigma_z \rangle)^2}.
\end{equation}
We shall now show how the dynamics can be approximated, up to first order in $\delta t$, with Eq. (5) of the main text. When a current sample $I$ is measured, the conditional evolution over a single time step is given by
\begin{equation}
\label{eq:matrix-formalism}
\rho(t+\delta t) = \frac{U_{t,t+\delta t} M_I \rho(t) M_I^{\dagger} U_{t,t+\delta t}^{\dagger}}{U_{t,t+\delta t} M_I \rho(t) M_I^{\dagger} U_{t,t+\delta t}^{\dagger}},
\end{equation}
where $M_I$ have been given in Eq. (6) of the main text. At first we are going to assume $H=0$, i.e. $U_{t,t+\delta t} = \mathbb{1}$. Adding the unitary term will be then straightforward. In the measurement basis, Eq. (\ref{eq:matrix-formalism}) can be written as 
\begin{equation}
\label{eq:component-formalism}
\begin{aligned}
\rho_{00}(t + \delta t) &= \frac{\rho_{00}(t) P_0(I)}{\rho_{00}(t) P_0(I)+\rho_{11}(t) P_1(I)},\\
\rho_{01}(t + \delta t) &= \frac{\rho_{01}(t) \sqrt{P_0(I)P_1(I)}}{\rho_{00}(t) P_0(I)+\rho_{11}(t) P_1(I)}.
\end{aligned}
\end{equation}
Notice that, for $\delta t^{-1} \gg \Gamma_d$ 
we have $P(I)  \simeq \rho_{00} P_0(I)+\rho_{11} P_1(I)$. Substituting this expression into Eq. (\ref{eq:component-formalism}) and using $\bar{\xi} = I- \sqrt{\Gamma_d} \langle \sigma_z \rangle$, we get
\begin{equation}
\label{eq:component-formalism-discrete}
\begin{aligned}
\rho_{00}(t + \delta t) &= \rho_{00}(t) \text{e}^{-\delta t \frac{(\bar{\xi} - q_1)^2}{2}} \text{e}^{\delta t \frac{\bar{\xi}^2}{2}},\\
\rho_{01}(t + \delta t) &= \rho_{01}(t) \sqrt{\text{e}^{-\delta t \frac{(\bar{\xi} - q_0)^2}{2}} \text{e}^{-\delta t \frac{(\bar{\xi} - q_1)^2}{2}}} \text{e}^{\delta t \frac{\bar{\xi}^2}{2}},
\end{aligned}
\end{equation}
where we have defined $q_i = -2 \sqrt{\Gamma_d} \rho_{ii}$. Expanding up to second order in $q_1$ and $q_2$ and approximating $\bar{\xi}^2 = \delta t$ (cf. Ref.~[\onlinecite{kb:214-wiseman-milburn}]) gives us Eq.~(\ref{eq:gambetta-component}) with $H=0$. Introducing now the unitary term through Eq. (\ref{eq:matrix-formalism}) and expanding up to the leading order in $\delta t$, one can show that the full structure of Eq.~(\ref{eq:gambetta-component}) is reproduced, thus justifying our model.

\end{document}